\documentclass[10pt,twocolumn,preprintnumbers,amsmath,amssymb,nofootinbib
,superscriptaddress]{revtex4-1}
\usepackage{graphicx,longtable,mathrsfs,color,array}
\usepackage[hidelinks]{hyperref}
\usepackage[usenames,dvipsnames]{xcolor} % colour
\usepackage{amssymb,amsmath,mathtools,mathrsfs,slashed} % maths
\usepackage{epsfig,subfigure,placeins,float} % plots
\usepackage{booktabs,longtable,ctable,multirow} % tables
\usepackage{exscale,relsize} % scale 
\usepackage[normalem]{ulem} % editing
\usepackage{enumerate}
\usepackage{times, mathptmx} % Times font
\usepackage[utf8]{inputenc}
\usepackage{color}
\usepackage{hyperref}
\usepackage{graphicx}% Include figure files
\usepackage{color}
\usepackage{graphicx,graphics}
\usepackage{bm}
\usepackage{tabularx}
\allowdisplaybreaks[1]

\begin{document}

\title{Scale Invariant Gravity and Black Hole Ringdown}
\author{Pedro G. Ferreira}
\email{pedro.ferreira@physics.ox.ac.uk}
\affiliation{Astrophysics, University of Oxford, DWB, Keble Road, Oxford OX1 3RH, UK}
\author{Oliver J. Tattersall}
\email{oliver.tattersall@physics.ox.ac.uk}
\affiliation{Astrophysics, University of Oxford, DWB, Keble Road, Oxford OX1 3RH, UK}
\date{Received \today; published -- 00, 0000}

\begin{abstract}
Scale invariant theories of gravity give a compelling explanation to the early and late time acceleration of the Universe. Unlike most scalar-tensor theories, fifth forces are absent and it would therefore seem impossible to distinguish scale invariant gravity from general relativity. We show that the ringdown of a Schwarschild-de Sitter black hole may have a set of massive modes which are characteristic of scale invariant gravity. In principle these new modes can be used to distinguish scale invariant gravity from general relativity. In practice, we discuss the obstacles to generating these new massive modes and their detectability with future gravitational wave experiments but also speculate on their role in Kerr black holes. \end{abstract}

\maketitle
%%%%%%%%%%%%%%%%%%%%%%%%%%%%%%%%%%%%%%%%%%%%%%%%%%%%%%%%%%%%%%
\section{Introduction}
One of the overarching quests of modern physics is to find fundamental symmetries of nature. These can be used to unify theories through simplified mathematical structures. Gauge theories and general covariance are two particularly fruitful examples. Another, intriguing, possibility is global scale invariance or Weyl invariance . Such a symmetry arises when the theory is invariant under the rescaling of the fundamental fields. For example if the theory consists of a metric, $g_{\mu\nu}$ and a scalar field, $\varphi$, then it will be invariant under a global transformation of the form $g_{\mu\nu}\rightarrow \lambda^2g_{\mu\nu}$ and $\varphi\rightarrow \lambda^{-1}\varphi$, where $\lambda$ is a constant. This symmetry can be made local, by gauging in a way which is entirely analogous to what one does with $U(1)$ in scalar electro-dynamics.

Over the past few years, scale invariant gravity has been extensively studied \cite{Ferreira:2016vsc,Ferreira:2016wem,GarciaBellido:2011de,Rubio:2014wta,Bezrukov:2014ipa,Karananas:2016grc,Wetterich:1987fm,Kallosh:2013maa,Carrasco:2015rva,Quiros:2014wda,Kurkov:2016zpd,Karananas:2017zrg,Karam:2017zno,Rubio:2017gty,Karananas:2017mxm,Kannike:2016wuy,Einhorn:2017icw,Salvio:2017qkx,Salvio:2017xul,Salvio:2014soa, Ghilencea:2018thl,Ghilencea:2018dqd,Ghilencea:2019jux} (although see \cite{Utiyama:1973nq,Utiyama:1974qu,Smolin:1979uz,Nishino:2007mj,Nishino:2009zza,Nishino:2011zz} for some earlier work). It has been shown to have a novel form of symmetry breaking -- inertial symmetry breaking -- in which scale emerges spontaneously without recourse to an explicitly symmetry breaking potential \cite{Ferreira:2018itt}. The symmetry broken state is an attractor of the dynamics and links the ultraviolet behaviour -- the effective Planck mass -- with the infrared behaviour -- the effective cosmological constant -- through ratios of dimensionless, fundamental constants. It has been shown that in certain, simple, scenarios, it is possible to obtain an inflationary period at early times as well as a late time period of accelerated expansion. While the radiative stability of such a construction can be problematic, the idea of {\it quantum} scale invariance as a fundamental principle and how it is incorporated in renormalization and regularization is a fruitful avenue of research.

From the gravitational point of view, scale invariance has been shown to lead to an intriguing phenomenon. Current implementations of scale invariance involve scalar tensor theories. It is well-known that scalar tensor theories lead to fifth forces which are tightly constrained both astronomically and in the laboratory. It has been shown, however, that in scale invariant gravity, these fifth forces are absent. In essence, the fifth force is mediated by the dilaton in the theory which, in the case of a scale invariant matter sector is completely decoupled \cite{Brax:2014baa,Ferreira:2016kxi}. Hence, scale invariant gravity evades fifth force constraints. It would seem, therefore, that it is impossible to identify an observable, gravitational, signature of scale invariance.

One arena where one might look for signatures of scale invariance is near black holes. A priori, such a regime might not look too promising. As mentioned above, scale invariance is implemented in scalar-tensor theories of gravity which have been shown to satsify variants of no-hair theorems \cite{Cardoso:2016ryw}. This means that black holes in such scale invariant theories are indistinguishable from those in general relativity (GR) -- Schwarzschild or, more generally Kerr-Newman. It is conventionally assumed that, if the black hole solutions of a modified theory gravity are indistinguishable from those of GR, then it is impossible to use them as laboratories or probes of new gravitational physics (although see \cite{Herdeiro:2015waa}).

It has been shown that, in fact, perturbations of Schwarzschild and Kerr-Newman black holes will carry information about extensions to general relativity \cite{Barausse:2008xv,Tattersall:2017erk,Tattersall:2018nve} . For example, in general scalar-tensor theories, there will be non-minimal coupling between the scalar and the metric sector. This means, even though the background scalar field (or fields) may be constant, perturbations in the scalar field will source perturbations in the metric and will, most notably affect the quasi-normal modes that emerge during the ringdown phase after black formation. A notable example of when this may happen is in the final phase of a binary black hole merger. Hence one might hope that a signature of scale invariance may be present in the quasi-normal mode spectrum of black holes in scale invariant gravity. In this paper we identify such signatures.

The structure of this paper is as follows. In Section \ref{SIG} we lay out a set of scale invariant theories involving one or multiple scalar fields. We describe their phenomenology and, in particular, how inertial symmetry breaking occurs. In Section \ref{QNM} we describe in some detail perturbations around a Schwarzschild black hole and how to determine associated quasi-normal modes (QNM) -- we emphasize here that we will focus on these modes and not, for example, quasi-bound states or other phenomena. We work our way through the case of one, two and then multiple scalar fields identifying the associated eigenmodes of the perturbation spectrum. Finally, in Section \ref{D} we discuss our results, and link it with previous findings about fifth forces and the dynamics of the dilaton in such theories.

%%%% SCALE INVARIANT GRAVITY
\section{Scale Invariant Gravity}
\label{SIG}
Let us begin with the simplest version of scalar tensor, scale invariant gravity:
\begin{eqnarray}
S=-\int d^4x\sqrt{-g}\left[\frac{1}{12}\alpha \phi^2  R + \frac{1}{2} \partial_\mu \phi \partial^ \mu \phi +\lambda \phi^4 \right],
\end{eqnarray}
where $g_{\mu\nu}$ is the space time metric and $\phi$ is the scalar field. This theory is invariant under $g_{\mu\nu}\rightarrow \lambda^2g_{\mu\nu}$ and $\phi\rightarrow \lambda^{-1}\phi$, where $\lambda$ is a constant. Note that the effective Planck mass is $M^2=-\alpha\phi^2/6$ and hence $\alpha$ should be negative. Note also that this theory is conformally invariant if $\alpha=1$ -- in that case we can promote $\lambda$ to a field, $\lambda(x^\mu)$.

The evolution equation for the scalar field can be rewritten in the form of a conserved current
\begin{eqnarray}
\nabla^\alpha K_\alpha=0 \label{k1}
\end{eqnarray}
where $\nabla^\alpha$ is the covariant derivative and
\begin{eqnarray}
K_\alpha=(1-\alpha)\phi\partial_\alpha \phi
\end{eqnarray}
Note that $K_\alpha$ can be expressed as
\begin{eqnarray}
K_\alpha=\partial_\alpha K \label{k2}
\end{eqnarray}
where the kernel, $K$ is given by
\begin{eqnarray}
K=\frac{1}{2}(1-\alpha)\phi^2
\end{eqnarray}
In a homogeneous, expanding, background, $g_{\alpha\beta}=(-1,a^2\delta_{ij})$ where $a$ grows with time, we have that $K\rightarrow K_0$ and scale invariance is spontaneously broken even though no explicit scale is introduced into the action; the final symmetry scale is a remnant of the initial conditions of the scalar field. The resulting non-scale invariant theory has
\begin{eqnarray}
K_0&=&\frac{1}{2}(1-\alpha)\phi_0^2 \nonumber \\
M^2_{\rm Pl}&=&-\frac{1}{6}\alpha\phi^2_0 \nonumber \\
\Lambda&=&-6\frac{\lambda}{\alpha}\phi_0^2
\end{eqnarray}
Thus, the ratios of all emergent scales in this theory are dependent on the fundamental, dimensionless constants in the action.

We can generalize this construction to multiple scalar fields. We then have
\begin{eqnarray}
S=-\int d^4 x\sqrt{-g}\left[\frac{1}{12}\sum_{i}^{N}\alpha_i\phi_i^2R+\frac{1}{2}\sum_{i}^{N}\partial_\mu\phi_i\partial^\mu\phi_i +W({\vec \phi})\right] \nonumber \\ \label{msaction}
\end{eqnarray}
where the potential has the form:
\begin{eqnarray}
W({\vec \phi})=\sum_{i}^N\sum_{i}^N \phi^2_i \lambda_{ij}\phi^2_j \nonumber
\end{eqnarray}
which a generalizes $\lambda\phi^4$ from the single field case. As in the case of the single field case, we can find a kernel
\begin{eqnarray}
K=\frac{1}{2}\sum_{i=1}^N(1-\alpha_i)\phi^2_i \label{ellipse}
\end{eqnarray}
which also evolves following equations \ref{k1} and \ref{k2}. Again, scale invariance is spontaneously broken but now the broken scale invariance phase lives an higher dimensional ellipsoide given by
\begin{eqnarray}
\sum_{i=1}^N(1-\alpha_i)\phi^2_i=2K_0
\end{eqnarray}
The direction connecting the origin to the ellipse is the dilaton, the Goldstone mode of the broken global symmetry, and decouples from the other degrees of freedom. 

While inertial symmetry pushes the fields onto the ellipsoid, there is, ultimately a fixed point. Minimizing the effective potential (which included the effect of the minimal coupling), one finds that the ratios of the fields are constrained by
\begin{eqnarray}
\sum_{jk}\phi^2_j{\cal A}^{(i)}_{jk}\phi^2_k=0
\end{eqnarray}
where
\begin{eqnarray}
{\cal A}^{(i)}_{jk}=\lambda_{jk}-\frac{\alpha_j}{\alpha_i}\lambda_{ik}
\end{eqnarray}
If all the $\alpha_i$ are different, the matrices ${\cal A}^{(i)}_{jk}$ have rank $N-1$ and the constraint corresponds to a line in field space. The intersection of this line with the ellipsoide give us the fixed point, or ground state of the system. As in the case of the single scalar field, we can then determine the effective Planck mass and cosmological constant from the values of the fields at this fixed point. It is around this vacuum that we will study black hole solutions in later sections.

It is useful to focus on the particular case of two scalars as it has been extensively studied before. In that case the symmetry broken phase lies on an ellipse given by
\begin{eqnarray}
(1-\alpha_1)\phi_1^2+(1-\alpha_2)\phi_2^2=2K_0\label{ellipseN2}
\end{eqnarray}
and has a fixed point at
\begin{eqnarray}
\frac{\phi_{1,0}^2}{\phi_{2,0}^2}=\frac{\alpha_1\lambda_{22}-\alpha_2\lambda_{12}}{\alpha_2\lambda_{11}-\alpha_1\lambda_{12}}\label{fixedpoint}
\end{eqnarray}

We have that, at the fixed point, we can determine the the effective cosmological constant in terms of the scalar field:
\begin{eqnarray}
\Lambda=6\frac{\lambda^2_{12}-\lambda_{11}\lambda_{22}}{\alpha_1\lambda_{22}-\alpha_2\lambda_{12}} \phi_{1,0}^2 \label{CCfixedpoint}
\end{eqnarray}
Note that, unlike in the case of the single scalar field, here we can choose the coupling constants such that $\Lambda=0$ while $\phi_{1,0}^2,\phi_{2,0}^2\neq0$.

The cosmological evolution of the 2-field case has been extensively studied. It can be shown that there exists an initial period of slow roll during which the universe inflates. If $\alpha_1<\alpha_2$, we have that $M^2_{\rm Pl}$ is initially primarily set by $\phi_2$. When $\alpha_1\phi^2_1=\alpha_2\phi^2_2$, the universe exits inflation and after a period of sub-luminal expansion, it ends up at the fixed point with stable Planck and cosmological constants. In this regime it would seem that the theory is indistinguishable from Einstein gravity.

It is well established that, in general, black-holes in scalar tensor theories are indistinguishable from those in Einstein gravity. This is normally stated as black holes in scalar-tensor theories having no hair \cite{Hui:2012qt,Sotiriou:2015pka,2014PhRvD..89h4056G}. There are a number of counterexamples (for example, coupling a scalar field to the Gauss-Bonnet invariant \cite{Sotiriou:2015pka}), violating assumptions that go into the no-hair theorems; an active field of research is to determine how dynamics and environment can lead to observable hair. Nevertheless, the particular models we look at here lie firmly in the region of theory space which satisfy the no-hair theorem. In practice, this means that black holes in these theories have the Schwarzschild-de Sitter backgrounds.

As mentioned, in the introduction, the fact that black holes have no hair in these theories does not mean it isn't possible to pick up a signature of the scalar field. We have fleshed out the idea that, even though the background is indistinguishable from GR, {\it perturbations} on that background might not be \cite{Barausse:2008xv}. The idea (which will be further developed in the next section) is that the extra degrees of freedom (i.e. the scalar fields) may be excited and through the non-minimal coupling to the metric sector will contaminate the gravitational wave sector. As a result, there will be a superposition of quasi-normal modes: the original, GR-like modes and the new, scalar field sourced, modes. We have argued that, under certain assumptions, these new modes may be detected with future gravitational wave experiments. 
Scale invariant theories give us a clear, worked out examples of how these new quasi-normal modes can emerge. In the following section we explore this happens

%%%%%%% Perturbations
\section{Perturbations}
\label{QNM}
%\subsection{Framework}

When considering perturbations to spherically symmetric spacetimes, it is natural to decompose perturbations using tensorial spherical harmonics \cite{1975RSPSA.343..289C,0264-9381-16-12-201,Kokkotas:1999bd,Berti:2009kk,Martel:2005ir}. One finds that perturbations possess a definite parity under inversion, odd (or axial) and even (or polar). Much like how scalar, vector, and tensor perturbations decouple from one another in cosmological perturbation theory, the odd and even parity perturbations decouple from one another when considering perturbations of spherically symmetric black holes. 

As the non-minimally coupled scalar fields present in scale invariant gravity are of even parity, the odd sector of the gravitational perturbations is completely unaffected. Thus we recover the GR result that the odd parity metric degree of freedom obeys the Regge Wheeler equation for a Schwarzschild-de Sitter black hole \cite{Regge:1957td,Cardoso:2001bb}. We will thus focus on even parity perturbations of the black hole and scalar(s) for the rest of this section. 

We first decompose the metric $g$ into the background Schwarzschild-de Sitter and a small perturbation $h$:
\begin{align}
g_{\mu\nu}=\;\overline{g}_{\mu\nu}+h_{\mu\nu}
\end{align}
such that the $\overline{g}$ is given by
\begin{align}
\overline{g}_{\mu\nu}dx^\mu dx^\nu=&\;-f(r)dt^2+f(r)^{-1}dr^2+r^2d\Omega^2\\
f(r)=&\;1-\frac{2M}{r}-\frac{\Lambda}{3}r^2
\end{align}
with $M$ being the black hole mass {(note that we are setting $G=c=1$ and so $M$ has units of length)} and $\Lambda$ is a (positive) effective cosmological constant. 

The even parity metric perturbation is then given by (in Regge-Wheeler gauge) \cite{Zerilli:1970se}:
\begin{align}
 h_{\mu\nu,\ell m}^{even}=& \begin{pmatrix}H_0(r)&H_1(r)&0&0\\H_1(r)&H_2(r)&0&0\\0&0&K(r)r^2&0\\0&0&0&K(r)r^2\sin\theta \end{pmatrix}Y^{\ell m}e^{-i\omega t}.
\end{align}
We further choose to decompose perturbations of any scalar fields $\phi_i$ such that:
\begin{align}
\phi_i=\phi_{i,0}\left(1+\frac{\varphi_i(r)}{r}Y^{\ell m}e^{-i\omega t} \right),
\end{align}
with $|\varphi_i|\ll\phi_{i,0}$. The $H_i$, $K$ and $\varphi_i$ are radial wavefunctions describing the perturbations, whilst the $Y^{\ell m}$ are the usual spherical harmonics. Due to the static nature of the background we've assumed a harmonic time dependence of $e^{-i\omega t}$. Note that we have suppressed spherical harmonic indices on the radial wavefunctions and the $\omega$; in general the perturbations of both the metric and scalar fields will be represented by a sum over $\ell$ of the modes (we will find that the perturbations are independent of $m$ due to the spherical symmetry of the problem, and thus we are free to set $m=0$).

Schematically, a field $\psi(r)$ propagating on a spherically symmetric black hole background obeys (in most cases) an equation of the form:
\begin{align}
\left[\frac{d^2}{dr_\ast^2}+\omega^2-V(\ell,r_\ast)\right]\psi(r_\ast)=\;S.\label{schroeq}
\end{align}
where $r_\ast$ is the tortoise coordinate defined by $dr=f(r)dr_\ast$, such that $-\infty<r_\ast<\infty$ from the black hole horizon to spatial infinity, and $S$ is some source term (which may or may not be zero depending on the details of the gravity theory). We see that this equation is a second-order Schr{\"o}dinger style wave equation, where the role of the `energy' is played by $\omega^2$.

One can show that, with boundary conditions such that the propagating field $\psi$ is purely outgoing at each boundary of the domain (i.e. with no waves originating from within the black hole horizon or from spatial infinity), the solutions to eq.~(\ref{schroeq}) lead to a discrete spectrum of complex frequencies $\omega$ for each value of $\ell$. Such frequencies are known as the Quasi-Normal Modes (QNMs) of the system, and they describe the oscillation and damping times of the exponentially damped sinusoidal waves emitted by each perturbed field. Thus at the end of a binary black hole merger, for example, when we are left with a highly perturbed remnant black hole, we expect to see this `ringdown' section of exponentially damped gravitational waves in the observed signal.

We note that we will focus solely on waves with these boundary conditions in this paper. There are, however, other alternatives; for example, sending the value of the wave to zero at infinity may lead to a different phenomenon -- quasibound states -- in which the scalar field accumulates around the horizon \cite{Detweiler:1980uk,Dolan:2007mj}; super-radiance \cite{Press:1972zz,Cardoso:2004nk,Brito:2015oca}, an instability  which emerges in the Kerr solution, may also be triggered. Furthermore, in de Sitter space there exists another family of modes associated with the cosmological horizon, and which is present even in the absence of a black hole (i.e. in the $M=0$ limit) \cite{Jansen:2017oag,Cardoso:2017soq,Dias:2018etb}. In this paper we choose not to look into these states and leave this to future work, instead focussing on the familiar `photon sphere' family of QNMs, which asymptote to the Schwarzschild QNMs in the case of a vanishingly small cosmological constant.

It is standard practice to use the fact that $H_i$ and $K$ can be expressed in terms of single field, $\Psi$ through
\begin{align}
\Psi=\frac{1}{3M+Lr}\left(K(r)r^2+\frac{rf(r)}{i\omega}H_1(r)\right),\label{psiGR}
\end{align}
with $2L=(\ell+2)(\ell-1)$, and where $H_0$ and $H_2$ are shown to be auxiliary fields through the Einstein equations. In the following sections we will see that the perturbed field equations of scale invariant gravity can be manipulated in such a way that $\Psi$ can be combined with the scalar degrees of freedom $\varphi_i$ into a single master variable ${\tilde \Psi}$ that obeys the Zerilli equation, the wave equation in the form of eq.~(\ref{schroeq}) that usually describes even parity metric perturbations in GR. The scalar fields, on the other hand, form a coupled system of similar style wave equations. 

Despite the master variable ${\tilde \Psi}$ obeying the same Zerilli equation as in GR, due to the presence of the scalar perturbations in the definition of ${\tilde \Psi}$, we will see that the scalar fields act as a source for the evolution of the metric variables, driving the gravitational field oscillations at characteristic frequencies associated with the \textit{scalar} QNM spectrum. This phenomenon has previously been observed in the case of Chern-Simons gravity, where the scalar perturbations are coupled to the \textit{odd} parity metric perturbation \cite{Molina:2010fb}. In this way, the gravitational wave emission from a Schwarzschild-de Sitter black hole in scale invariant gravity can be modified from that expected in GR due to excitation of the scalar perturbations, due to the emission of gravitational wave at the `transient' frequencies associated with the regular GR QNM spectrum, and those oscillating at the forced scalar frequencies.

%%%SINGLE SCALAR FIELD CASE
\subsection{Single scalar field}

Consider first the case of a single scalar field which is on the fixed point. In this symmetry broken phase, we need study perturbations around  a Schwarzschild-de Sitter background with no non-trivial scalar hair but with an effective cosmological constant $\Lambda$ is given by:
\begin{align}
\Lambda=-6\frac{\lambda}{\alpha}\phi_0^2.
\end{align}
We have, then, for the even parity sector of the perturbations:
\begin{align}
\left[\frac{d^2}{dr_\ast^2}+\omega^2-V_Z(r)\right]{\tilde \Psi}(r)=&\;0\label{eqZN1}\\
\left[\frac{d^2}{dr_\ast^2}+\omega^2-U_0(r)\right]\varphi(r)=&\;0
\end{align}
where ${\tilde \Psi}$ is related to the original `metric only' field $\Psi$ (given by eq.~(\ref{psiGR})) through:
\begin{align}
{\tilde \Psi}=\Psi+\frac{2r}{3M+Lr}\varphi(r).
\end{align}
So, even though the system is decoupled into a scalar field perturbation $\varphi$  equation and a combined metric-scalar master variable ${\tilde \Psi}$ that obeys the standard GR Zerilli equation for a Schwarzschild-de Sitter background, one can see the explicit sourcing of the pure metric perturbations, $\Psi$ by the scalar perturbation $\varphi$ in eq.~(\ref{eqZN1}).

The potentials are
\begin{align}
V_Z(r)=&\;\frac{2f(r)}{r^3(3M+L r)^2}\left(9M^3+3L^2Mr^2+L^2(1+L)r^3\right.\nonumber\\
&\left.+3M^2(3L r-r^3\Lambda)\right)\\
U_{\mu^2}(r)=&\;f(r)\left(\frac{\ell(\ell+1)}{r^2}+\frac{f'(r)}{r}+\mu^2\right)
\end{align}
We can see then a feature which was emphasized in \cite{Tattersall:2019pvx}: as well as the usual general relativistic modes that arise from the Zerilli equation, there will be a new set of modes, injected by the scalar field perturbations. 

In this particular case, the additional modes are those of a massless scalar field (i.e. $\mu^2=0$) which is to be expected. The scalar field plays the role of the dilaton and, in the symmetry broken phase, is a massless Goldstone boson. Hence, we are seeing the imprint of the broken global symmetry on the equations and resulting QNMs.

Assuming a small $\Lambda$, we can compare the $\ell=2$ modes from the gravitational and scalar spectra (to 3 significant figures) using the results of \cite{Tattersall:2018axd}:
\begin{align}
M\omega_g=&\; 0.374-0.0887i+\Lambda M^2\left(-1.67+0.333i\right)+\mathcal{O}(\Lambda^2M^4)\\
M\omega_s=&\;0.484-0.0968i+\Lambda M^2\left(-2.35+0.373i\right)+\mathcal{O}(\Lambda^2M^4).
\end{align}
More accurate analytic expressions for the gravitational and scalar frequencies as a function of $\Lambda$ can be found in \cite{Tattersall:2018axd}, whilst the QNMs for a variety of fields on a Schwarzschild-de Sitter background were calculated using 6th order WKB methods in \cite{Zhidenko:2003wq}. Note however, that corrections to the Schwarzschild QNM spectra due to $\Lambda$ are negligible: for Solar mass black holes we have $\Lambda M^2\sim 10^{-46}$ while for supermassive black holes (which can be up to $10^9$ time more massive) we have $\Lambda M^2\sim 10^{-28}$. 

%%TWP FIELDS
\subsection{Two field model}
We now consider the two field case which has more involved dynamics. Recall that the symmetry broken phase lies on an ellipse and the end point is a fixed point; it is possible, however, to remain in the same vacuum by moving along the ellipse.
For even parity perturbations, we again find that a master variable ${\tilde \Psi}$ obeys the Zerilli equation as in eq.~(\ref{eqZN1})
%\begin{align}
%\left[\frac{d^2}{dr_\ast^2}+\omega^2-V_Z\right]{\tilde \Psi}(r)=&\;0\label{eqZN2},
%\end{align}
and is now is given by:
\begin{widetext}
\begin{align}
{\tilde \Psi}=&\;\Psi+\frac{2r}{3M+Lr}\frac{\varphi_1(r)\alpha_1(\alpha_1\lambda_{22}-\alpha_2\lambda_{12})+\varphi_2(r)\alpha_2(\alpha_2\lambda_{11}-\alpha_1\lambda_{12})}{\alpha_1^2\lambda_{22}+\alpha_2^2\lambda_{11}-2\alpha_1\alpha_2\lambda_{12}}.\label{eqPsiN2}
\end{align}
\end{widetext}
As for the single field case, if we were to again split ${\tilde \Psi}$ into the pure metric field $\Psi$ and scalar contributions, eq.~(\ref{eqZN1}) would show $\Psi$ being sourced by both $\varphi_1$ and $\varphi_2$. 

The perturbations from the two scalar fields form the following coupled system of equations:
\begin{align}
\frac{d^2}{dr_\ast^2}\begin{pmatrix}\varphi_1 \\ \varphi_2 \end{pmatrix}=\;\bm{U}\begin{pmatrix}\varphi_1 \\ \varphi_2 \end{pmatrix}
%=\;\begin{pmatrix}U_{11}(r) & U_{12}(r) \\U_{21}(r) & V_{22}(r)\end{pmatrix}\begin{pmatrix}\varphi_1 \\ \varphi_2 \end{pmatrix}
\label{phieqN2}
\end{align}
where the the potential matrix $\bm{U}$ is given by:
\begin{align}
\bm{U}=\;-\begin{pmatrix}\omega^2-U_{\mu_1^2}(r)&f(r)\mu_{1}^2\\f(r)\mu_{2}^2&\omega^2-U_{\mu_1^2}(r)\end{pmatrix}
\end{align}
and the effective masses $\mu_i$ are given by:
\begin{widetext}
\begin{align}
\mu_1^2 = &\; \frac{8(\alpha_2-1)(\alpha_1\lambda_{12} - \alpha_2\lambda_{11})^2}{(\alpha_1-1)\alpha_1^2\lambda_{22} + (\alpha_2-1)\alpha_2^2\lambda_{11} - \alpha_1\alpha_2\lambda_{12}(\alpha_1+\alpha_2-2)}\phi_{1,0}^2\\
\mu_2^2 =&\; \frac{8(\alpha_1-1)(\alpha_2\lambda_{12} - \alpha_1\lambda_{22})^2}{(\alpha_1-1)\alpha_1^2\lambda_{22} + (\alpha_2-1)\alpha_2^2\lambda_{11} - \alpha_1\alpha_2\lambda_{12}(\alpha_1+\alpha_2-2)}\phi_{2,0}^2\label{phimasses}
\end{align}
\end{widetext}

Despite $\bm{U}$ having non-constant components, we find that its eigenvectors \textit{are} constant. Defining a matrix of (column) eigenvectors $\bm{T}$:
\begin{align}
\bm{T}=\;\begin{pmatrix}1&-\mu_1^2/\mu_2^2\\ 1 & 1 \end{pmatrix}
\end{align}
and making a field redefinition such that:
\begin{align}
\begin{pmatrix}\varphi_1 \\ \varphi_2 \end{pmatrix}=\;\bm{T}\begin{pmatrix}\vartheta_1 \\ \vartheta_2 \end{pmatrix}
\end{align}
we find that the fields $\vartheta_1$ and $\vartheta_2$ decouple to form the following system of equations:
\begin{align}
\frac{d^2}{dr_\ast^2}\begin{pmatrix}\vartheta_1 \\ \vartheta_2 \end{pmatrix}&=\;\bm{T^{-1}UT}\begin{pmatrix}\vartheta \\ \upsilon \end{pmatrix} =\;\bm{\overline{\bm{U}}}\begin{pmatrix}\vartheta_1 \\ \vartheta_2\end{pmatrix}
\end{align}
where $\overline{U}_{12}=\overline{U}_{21}=0$ and:
\begin{subequations}
\begin{align}
-\overline{U}_{11}=&\;\omega^2-U_0(r)\\
-\overline{U}_{22}=&\;\omega^2-U_{\mu_+^2}(r).
\end{align}
\end{subequations}
where $\mu_+^2=\mu_1^2+\mu_2^2$.

We see that the coupled system of the massive scalar fields $\varphi_1$ and $\varphi_2$ is equivalent to a decoupled system of a massless scalar field $\vartheta_1$ and a massive scalar field $\vartheta_2$, such that the effective mass of $\vartheta_2$ is equal to the sum in quadrature of the individual effective masses of $\varphi_1$ and $\varphi_2$.

If we now consider the `full', rather than fractional, scalar perturbations $\delta\phi_i=\phi_{i,0}\varphi_i$ where the background values of the scalar fields are given by eq.~(\ref{fixedpoint}), we find the following matrix of (column) eigenvectors $\bm{\overline{\bm{T}}}$ in the $\delta\phi_i$ basis:
\begin{align}
\bm{\overline{\bm{T}}}=&\;\begin{pmatrix}\phi_{1,0}&0\\0&\phi_{2,0}\end{pmatrix}\begin{pmatrix}1&-\mu_1^2/\mu_2^2\\ 1 & 1 \end{pmatrix}\nonumber\\
=&\;\frac{1}{\phi_{2,0}}\begin{pmatrix}\phi_{1,0}/\phi_{2,0} & -\mu_1^2\phi_{1,0}/\mu_2^2\phi_{2,0}\\1&1\end{pmatrix}.
\end{align}
Using the background values of the scalar fields given by eq.~(\ref{fixedpoint}) and the expressions for the $\mu_i$ given by eq.~(\ref{phimasses}), the new (non-normalised) eigenvector corresponding to the massive mode can be shown to be:
\begin{align}
e_{massive}=&\;\begin{pmatrix}-\frac{\alpha_2-1}{\alpha_1-1}\sqrt{\frac{\alpha_1\lambda_{12}-\alpha_2\lambda_{11}}{\alpha_2\lambda_{12}-\alpha_1\lambda_{22}}}\\1\end{pmatrix}.
\end{align}
This is nothing more than the tangent vector to the ellipse in scalar field space defined in eq.~(\ref{ellipse}). 

The massless eigenmode, on the other hand, is in the direction:
\begin{align}
e_{massless}=\begin{pmatrix}\sqrt{\frac{\alpha_2\lambda_{12}-\alpha_1\lambda_{22}}{\alpha_1\lambda_{12}-\alpha_2\lambda_{11}}}\\1\end{pmatrix}.
\end{align}
Again, this is the goldstone mode arising from the breaking of the global symmetry, just as we saw in the case of the single scalar field. If we look at an analogous situations -- that of perturbations arising in inflation in this model -- these massless modes correspond to the isocurvature fluctuations in the primordial universe.

An interesting point to note is that the potential matrix $\bm{U}$ is not in general symmetric in either the $\delta\phi_i$ or $\varphi_i$ basis -- the frequencies are complex -- so we do not expect the eigenvectors to be mutually orthogonal. This can be understood geometrically: the massless mode is in the direction of the position vector of the fixed point while the massive direction is along the tangent. Only in the case of a circle ($\alpha_1=\alpha_2$) do we have that these two directions are orthogonal -- in general they are not.

Including first order corrections in $\mu_+^2$ and $\Lambda$ to the scalar frequencies, we find the following expression for the $\ell=2$ scalar mode:
\begin{align}
M\omega_s=\;0.484&-0.0968i+\Lambda M^2\left(-2.35+0.373i\right)\nonumber\\
&+(\mu_+M)^2\left(0.317+0.108i\right)+\mathcal{O}(\Lambda^2M^4). \label{QNMW2}
\end{align}
We can see the corrections due to $\Lambda$  are negligible but there is now a potentially non-trivial correction due to the massive mode. We will discuss this correction in more detail in the Section \ref{D}. 

%%STAROBINSKY
\subsection{Scale invariant Starobinsky model}

A particularly interesting case  arises if we consider a scale invariant version of the Starobinsky inflationary action \cite{Starobinsky:1980te,Ferreira:2019zzx} (but see also \cite{Rinaldi:2015uvu,Bamba:2015uxa,Tambalo:2016eqr,Karam:2018mft,Kubo:2018kho,Herrera:1995me,Maeda:1987xf,Wang:1990pr,Wetterich:2019qzx}),
%\begin{widetext}
\begin{align}
S=-\int d^4x\sqrt{-g}&\left[\frac{1}{12}\alpha_1 \phi_1^2  R + \frac{1}{6f_0^2} R^2+\frac{1}{2} \nabla_\mu \phi_1 \nabla^ \mu \phi_1 \right.\nonumber\\
&\left.+\lambda_1 \phi_1^4 \right],\label{Snokin}
\end{align}
%\end{widetext}
This action can be recast as a two field field scale invariant model if we introduce $\phi_2$ as an auxiliary field. We then have a  model with no canonical kinetic term for $\phi_2$, such that the action is given by
\begin{widetext}
\begin{align}
S=-\int d^4x\sqrt{-g}\left[\frac{1}{12}\left(\alpha_1 \phi_1^2 + \alpha_2 \phi_2^2\right) R + \frac{1}{2} \nabla_\mu \phi_1 \nabla^ \mu \phi_1 +\lambda_{11} \phi_1^4 + \lambda_{22} \phi_2 ^4 \right],\label{Snokin}
\end{align}
\end{widetext}
and $\lambda_{22}=\frac{3}{2}f^2_0\left(\frac{\alpha_2}{12}\right)^2$.
We find that $\phi_2$ can still have dynamics through its coupling to $R$. For this model we find the same background solutions for the $\phi_i$ as in the ``normal" two field model, whilst eq.~(\ref{eqZN1}) and eq.~(\ref{eqPsiN2}) again hold for the metric perturbations.

For the scalar field perturbations, with background values again given by eq.~(\ref{fixedpoint}), the system of equations given by eq.~(\ref{phieqN2}) again holds with the same potential matrix, only now the effective masses are given by:
\begin{subequations}
\begin{align}
\mu_{1}^2 = &\;8\lambda_{11}\phi_{1,0}^2 \\
\mu_{2}^2 = &\;-4\frac{(\alpha_1-1)(\alpha_1\lambda_{22})^2}{\alpha_2^3\lambda_{11}}\phi_{2,0}^2 \label{starobinskymasses}
\end{align}
\end{subequations}

As the schematic form of the coupled system of scalar equations has not changed, this system is also clearly diagonalisable into a massive mode and a massless mode. Due to the effective masses changing, however, we find that the massive eigenmode in the $\delta\phi_i$ basis is modified in this model (the massless eigenmode, which is independent of the effective masses, is unchanged):
\begin{align}
e_{massive}=\;\begin{pmatrix}-\frac{\alpha_2}{\alpha_1-1}\sqrt{\frac{\alpha_2\lambda_{11}}{\alpha_1\lambda_{22}}}\\1\end{pmatrix}.
\end{align}

If we modify eq.~(\ref{ellipseN2}) to reflect the lack of canonical kinetic term for $\phi_2$, such that
\begin{align}
(1-\alpha_1)\phi_{1,0}^2-\alpha_2\phi_{2,0}^2=&\;2K_0
\end{align}
then the massive eigenmode again lies tangential to this ellipse.

%%MULTIPLE FIELDS
\subsection{$N$ scalar fields}

We now assume $N$ conformally coupled scalar fields but consider no explicit cross-couplings  between the scalars in the action, giving the following:
\begin{align}
S=-\int d^4x\sqrt{-g}\left[\sum_{i=1}^N \frac{1}{12}\left(\alpha_i\phi_i^2\right)\, R +  \frac{1}{2} \nabla_\mu\phi_i\nabla^\mu\phi_i+\lambda_i\phi_i^4\right].
\end{align}
where we define $\lambda_{ij}\equiv\lambda_i\delta_{ij}$ with no summation assumed on the right hand side.

We find the following pattern for the system of equations for the scalar perturbations $\varphi_i$:
\begin{align}
\frac{d^2\varphi_i}{dr_\ast^2}+\left[\omega^2-U_{\mu_i^2}\right]\varphi_i=&\;f(r)\alpha_i\sum_{j\neq i}\varphi_j c_j
\end{align}
where
\begin{align}
c_i=&\; \frac{4\Lambda}{3}\left[\displaystyle\alpha_i^2(\alpha_i-1)\prod_{k\neq i}\lambda_k\right]\left[\displaystyle\sum_{j=1}^N\alpha_j^2(\alpha_j-1)\prod_{k\neq j}\lambda_k\right]^{-1}\\
\mu_i^2=&\;-\alpha_i\sum_{j\neq i}c_j
\end{align}
This system is diagonalisable, leading to one massless scalar mode, and $N-1$ massive modes with effective masses $m_i^2$, such that:
\begin{align}
\sum_{i=1}^N\mu_i^2=\sum_{i=1}^{N-1}m_i^2.
\end{align}

For the metric perturbations, we find that the master variable $\tilde{\Psi}$ that satisfies the usual Zerilli equation is given by:
\begin{align}
\tilde{\Psi}=&\;\Psi+\frac{2r}{3M+Lr}\left[\displaystyle\sum_{i=1}^N\alpha_i^2\varphi_i\prod_{k\neq i}\lambda_k\right]\left[\displaystyle\sum_{i=1}^N\alpha_i^2\prod_{k\neq i}\lambda_k\right]^{-1}.
\end{align}
We see that, as before, the pure metric perturbations will be sourced by the scalar fields, acting as $N$ driving oscillators. 

We have checked that these patterns holds for $N=2,3$, and we expect that they should continue to hold for general $N$ conformally coupled scalar fields with scale invariant potentials and no explicit scalar-scalar interactions.

%%%DISCUSSION
\section{Discussion}
\label{D}
Scale-invariant gravity has a number of features: it leads to observationally consistent cosmological models, it can evade stringent fifth force constraints and it has attractive quantum properties from the point of view of naturalness. The question then arises: is there a distinct observational signature which we can look for in the current or future  experimental and astronomical endeavours?

We have chosen to look at black holes and gravitational waves. Specifically, we have focused on the ring down phase of perturbed Scwarzschild-de Sitter black holes (although our findings should be generalisable to Kerr geometries). \'A priori one might think this is a lost cause: scale invariant gravity is a scalar tensor theory which is known to obey variants of the no-hair theorem. In other words, black holes in such theories should be indistinguishable from those in general relativity. Thus we might not expect any distinctive signals  marking the presence of scale-invariance. 

We have recently shown, however, that, even in the case of hairless black holes, the non-minimal coupling between the scalar fields and the metric leads to new modes in the gravitational wave spectrum of perturbations. In this paper we have identified these modes for a range of scale invariant models. Unsurprisingly, in the simplest case of one scalar field,  this mode is the goldstone boson of the broken global scale invariance -- the dilaton; it has a massless spectrum, as one would expect. We also know that the dilaton is completely decoupled from the standard model if the universe is fully scale invariant. This means that it is impossible to excite these modes to begin with: the dilaton equations of motion are completely unsourced by matter. Hence this new mode will not be present the ringdown of a black hole merger event.

The situation becomes far more interesting if we consider two or more scalar fields. As has been shown previously, the vacuum state of the system is on a fixed point which lies on an ellipse where global scale invariance has been broken. Now there are two new modes, on top of the GR modes: the massless mode (the decoupled dilaton) and a new massive mode. The massive mode is tangent to the ellipse and is coupled to the rest of the universe. That is, it should be possible to excite it in the merger of compact objects (or neutron stars). Extending the scale invariance to more scalar fields will lead to more extra modes but, ultimately, the signature will be the same: a new, quasi-normal, mode which will coexist with the normal GR spectrum.

It is not enough to say that a new mode exists -- we need to be able to generate it. Unlike in the case of the dilaton, nothing stops the new mode from being generated during a merger or any other, violent astrophysical event. In practice, and as mentioned above, these theories satisfy no-hair theorems. Consider then
the merger of two black holes: during the inspiral, these black holes will be hairless and, unless there is some non-trivial dynamics during the merger, there is no way to generate a non-amplitude of the new mode that emerges during ringdown. If one of the objects involved in the binary is a neutron star, the situation is more promising. Furthermore, one can imagine the mergers are complex, dirty events, immersed in time varying cosmological backgrounds. These complexities break the conditions of the no-hair theorems and may lead to non-negligible scalar modes being excited and seeding the new massless mode. This non-dynamical generation of hair is an open question and the subject of further investigation \cite{Clough:2019jpm,Hui:2019aqm}.

The new quasi normal mode depends on the effective mass $\mu^2_+$ which we  will depend on the coupling constants of the theory and the symmetry breaking scale. We will now discuss three different regimes for the mass: the massless limit (or when $\mu_+\simeq \Lambda$), the intermediate limit (when $\mu_+ M\sim 1$) and the very massive limit, ($\mu\simeq \mu_{EW}$ or other high energy scales).

If we assume that $\mu^2_+\sim\Lambda$, the new mode is effectively massless and we have a clear prediction. To assess if it is observable, we first note that the  quasi-normal frequency of the (effectively) massless new mode is comparable (in magnitude) to the dominant quasi normal mode frequency from the normal, GR, spectrum. Let us then assume that the mode is generated with some initial amplitude $A_s$. To find out if it is detectable, we use the Fisher matrix analysis of \cite{Berti:2005ys,Tattersall:2019pvx}, and assume a gravitational waveform consisting of a superposition of $\ell=2$ modes from the gravitational and massless scalar spectra (assuming that the massless scalar mode is not excited). We find the following leading order requirement on the ratio of the amplitudes of the scalar ($A_s$) and gravitational modes ($A_g$) in order to resolve distinct frequencies and damping times in the gravitational wave signal:
\begin{align}
\frac{A_s}{A_g}\gtrapprox\frac{21}{\rho}
\end{align}
where $\rho$ is the total signal to noise ratio (SNR) of the gravitational wave signal, and we are assuming that the scalar amplitude will be subdominant compared to the gravitational amplitude. With a SNR of $\rho\sim 10^2$, which is believed to be eminently achievable with LISA, third generation ground based detectors, or through stacking several signals together \cite{Berti:2016lat,Yang:2017zxs,Yang:2017xlf,DaSilvaCosta:2017njq,2018PhRvD..98h4038B,2019arXiv190209199B}, a scalar amplitude of around tens of percent the strength of the gravitational mode would be required to discern the presence of a second mode. 

One should of course be mindful that this analysis assumes a simple two mode waveform model, with the `fundamental' $\ell=2$ mode from each of the gravitational and scalar spectra. To more accurately model ringdown and extract parameters from a gravitational wave signal, one should take into account higher overtones as well as fundamental modes \cite{Giesler:2019uxc}, and be aware that in some cases the amplitudes of `less dominant' modes (e.g. those with higher harmonic index $\ell$) may be comparable to the dominant modes \cite{Berti:2006wq,Kamaretsos:2011um,Zhang:2013ksa,London:2014cma}.  

If $\mu M\simeq 1$, we can use the forecasts we presented above but there are some qualitatively interesting aspects that we should highlight. The real part QNM frequency will grow with $\mu^2_+$ but depends very weakly on it. So large changes in $\mu^2_+$ leading to small changes in $\omega_R$. More interestingly \cite{1992CQGra...9..963S}, the imaginary part of the QNM frequency {\it decreases} with increasing $\mu^2_+$ which means that  the decay time is longer and thus these modes may be, marginally, more detectable.

A different regime is that $\mu^2_+$ is in fact, quite large. To understand why this is so, we need to remind ourselves that the threshold for whether the mode is massive or massless is (replacing dimensionful constants) set by $(G M_\odot/c^2)^{-1}\sim 10^{-10}$ eV. Let us then see what kind of masses we should expect in, for example, the Dilaton-Higgs model \cite{GarciaBellido:2011de,Rubio:2014wta}.
Simplifying the analysis by assuming $\Lambda\simeq0$ we have then  that the fixed point is
\begin{eqnarray}
\left(\frac{\phi_{1,0}}{\phi_{2,0}}\right)^2\simeq-\frac{\lambda_{12}}{\lambda_{11}}
\end{eqnarray} 
(note that $\lambda_{11}\lambda_{22}-\lambda_{12}^2\simeq0$). If we assign to $\phi_2$ the role of the Higgs \cite{GarciaBellido:2011de,Rubio:2014wta}, we have that its mass is given by
\begin{eqnarray}
\frac{m_H^2}{M^2_{\rm Pl}}\simeq
48\lambda_{12}\frac{(1-\alpha_1)-(1-\alpha_2)\frac{\lambda_{12}}{4\lambda_{22}}}{\alpha_1(1-\alpha_1)-\alpha_2(1-\alpha_2)\frac{\lambda_{12}}{4\lambda_{22}}}+{\cal O}(\lambda_{11})
\end{eqnarray}
We now have that the Higgs self-coupling satisfies $\lambda_{22}\sim 1$ while the rest of the potential couplings satisfy a hierarchy $\lambda_{11} \ll \lambda_{12} \ll \lambda_{22}$. Furthermore, from cosmological constraints \cite{Ferreira:2018qss} we have that $|\alpha_1|< 0.019$ and $\alpha_2< -0.048$. If we saturate the second bound, we can simplify both our expressions for $\mu_+^2$ and $m^2_H$ and we find that 
\begin{eqnarray}
\mu_+^2\simeq \frac{48\lambda_{12}}{\alpha_1}M^2_{\rm Pl}\simeq m^2_H \gg (10^{-10} \  {\rm eV})^2
\end{eqnarray}

Alternatively one can estimate the magnitude of radiative corrections to $\phi_1$ from its non-minimal coupling, generated at one loop between $\phi_1$ and $\phi_2$ in the scale-invariant Starobinsky model, this is given by
\begin{eqnarray}
\delta m^2_1=\frac{1}{4\pi}\alpha_1\alpha_2 f^4_0
\end{eqnarray} 
This coupling give a mass-squared correction to the massive mode, of order $\delta\mu^2_+\sim\alpha_2 M^2_{\rm Pl} f^4_0$. If this mode is to be observable in the ring down, it should be of order $\delta\mu^2_+< (10^{-38})^2 M^2_{\rm Pl}$. This allows us to place an upper limit on $f_0$
\begin{eqnarray}
f_0< \frac{6}{\alpha_1\alpha_2}\times 10^{-10} \ {\rm GeV}
\end{eqnarray} 
With this value of $f_0$, the amplitude of density perturbations would be far too small for the model to be viable cosmologically. 

The behaviour of black hole perturbations for large masses is more exotic. For a start, a WKB analysis \cite{1992CQGra...9..963S} shows that for QNMs to exist, there is an upperbound on the mass set by the maximum of the Zerilli potential. For small $\ell$ this is of order $\mu_+ M\lesssim 1$ but in the eikonal limit, it is $\mu_+ M\lesssim \ell/4$. This means that, for the high masses we are considering here, only the very high $\ell$ modes will by QNM. 

Additionally, for massive fields, new phenoma have to be taken into account. As mentioned in Section \ref{QNM}, there are alternative modes: quasi-bound states  and super-radiance. The latter which may appear in the case of a rotating black hole leading to what has been dubbed a 'black-hole bomb" \cite{Brito:2015oca}. This goes beyond the spherically symmetric, perturbative calculation we have undertaken here but certainly merits further analysis.

While it seemed that any signature of the scale-invariant is experimentally illusive, we have shown that it may, in principle, be possible to distinguish scale-invariant gravity from GR through black hole spectroscopy. For a particularly extreme choice of parameters, the signature is somewhat generic: a new massless mode in the QNM spectrum. Such a mode will arise in any non-minimally coupled scalar-tensor theory where the effective mass of the scalar is negligible. But, as we know, such fields have long range fifth forces which, generically, couple to matter and are strongly constrained by laboratory and astronomical experiments. So, if one were to find such a QNM yet no evidence of a new fifth scalar force, one might be inclined to consider the possibility that gravity is scale invariant. 

More generally, and within the context of most of the scale invariant models that have been proposed, the new QNM mode will be too massive to be detected. Its frequency (and as a result its decay time) will be far too high for it to be observed in current and future gravitational wave experiments. Instead, in that regime, instabilities may emerge which can be a signature of the extra, massive degree of freedom tied to scale-invariant gravity.

\textit{Acknowledgements ---} We are extremely grateful for discussions with V. Cardoso, K. Clough, C. Hill, M. Lagos and G. Ross. PGF acknowledges support from 
Leverhulme, STFC, BIPAC. This  project  has  received  funding  from  the  European  Re-search  Council  (ERC)  under  the  European  Union's Horizon  2020  research  and  innovation  programme  (grant  agreement No 693024). OJT acknowledges support from STFC. Part of this work was done at Fermilab, 
operated by Fermi Research Alliance, LLC under Contract No. DE-AC02-07CH11359 with the United States Department of Energy. 

\bibliographystyle{apsrev4-1}
\bibliography{InfSI}

\end{document}